**Polychromatic Excitation of Delocalized Long-Lived Proton Spin States in Aliphatic Chains**

Anna Sonnefeld, Geoffrey Bodenhausen, Kirill Sheberstov*

*Laboratoire des biomolécules, LBM, Département de chimie, École normale supérieure, PSL University, Sorbonne Université, CNRS, 75005 Paris, France*

07 July 2022

*Corresponding author

kirill.sheberstov@ens.psl.eu

## Abstract

Long-lived states (LLS) involving pairs of magnetically inequivalent but chemically equivalent proton spins in aliphatic $(CH_2)_n$ chains can be excited by simultaneous application of weak selective radio-frequency (RF) fields at $n$ chemical shifts by polychromatic spin-lock induced crossing (poly-SLIC). The LLS are delocalized throughout the aliphatic chain, by mixing of intrapair singlet states and by excitation of LLS comprising products of four or six spins. The measured lifetimes $T_{LLS}$ in a model compound are about 5 times longer than $T_1$, and are strongly affected by interactions with macromolecules.

## Key words

The discovery of long-lived states (LLS) in nuclear magnetic resonance (NMR) has opened important perspectives [1–10]. If the lifetimes $T_{LLS}$ are longer than the longitudinal relaxation times $T_1$, one can store nuclear spin hyperpolarization [11], observe hyperpolarized metabolites in magnetic resonance imaging (MRI) [12], probe slow chemical exchange [13], determine slow diffusion rates [14,15], and detect weak interactions between potential drugs and target proteins [16–18]. The methods were further developed in order to excite long-lived coherences (LLCs) that can have lifetimes longer than both $T_2$ and $T_1$ [19–21].

In a pair of two spins A and A', the imbalance between the population of the singlet state $p(S_0^{AA'})$ and the mean population of the triplet states $\langle p(T^{AA'}) \rangle = \frac{1}{3} \big( p(T_{+1}^{AA'}) + p(T_0^{AA'}) + p(T_{-1}^{AA'}) \big)$ is known as triplet-singlet population imbalance, which is immune against relaxation driven by in-pair dipole-dipole couplings [18]. Such an LLS can be described by a scalar product $\hat{I}^A \cdot \hat{I}^{A'}$, where $\hat{I}^p = \hat{I}_x^p + \hat{I}_y^p + \hat{I}_z^p, p \in \{A, A'\}$. To excite such an LLS, the two spins should have either different resonance frequencies (chemical inequivalence), or have different interactions with other spins (magnetic inequivalence) [24,25], or the high temperature approximation must be violated [26,27]. In systems with more than two spins, one can excite long-lived imbalances between states that belong to different symmetries of the spin permutation group [6,8,10,28–32].

In molecules that contain nearly equivalent spins, either with slightly different chemical shifts, or with slightly different scalar couplings to neighbouring spins, a two-spin LLS is almost an eigenstate and therefore does not require any *RF* fields to be sustained [4,21,33]. Such molecules often require challenging chemical synthesis of isotopically enriched molecules with near-equivalent $^{13}C$ [6,7,33,34] or $^{15}N$ [10,35,36] spin pairs, although molecules that contain nearly equivalent proton spins have also been studied [4,37]. Here we show that LLS can be readily excited and observed in $CH_2$ groups of common aliphatic chains, which are ubiquitous in chemistry.

In achiral molecules that contain at least two neighbouring $CH_2$ groups, the two protons of each $CH_2$ group are chemically equivalent (i.e., have the same chemical shifts) but they are generally magnetically inequivalent (i.e., have different scalar couplings to neighbouring $CH_2$ groups) [38]. Provided the substituents R and R' in a molecule $R(CH_2)_nR'$ are distinct, neighboring $CH_2$ subunits are often weakly coupled to each other in high magnetic fields, i.e., $J$-couplings between them are much smaller than difference between their chemical shifts. In a chain with $n = 3$ neighboring weakly coupled $CH_2$ groups, the spin system of the 6 protons can then be denoted by $AA'MM'XX'$ [39]. A good approximation of the 64 eigenstates of the Hamiltonian in this case can be constructed by direct products of the intra-pair singlet and triplet states. States that contain an even number of singlets, e.g., $S_0^{AA'}S_0^{MM'}T_i^{XX'}$, are globally *gerade* (*g*) with respect to *simultaneous* intra-pair permutation of spins $(AA')(MM')(XX')$ ,while products containing an odd number of singlets, like $S_0^{AA'}T_i^{MM'}T_j^{XX'}$ and $S_0^{AA'}S_0^{MM'}S_0^{XX'}$ are globally *ungerade* (*u*). The states can be separated into global (*g*) or (*u*) manifolds and further sorted with respect to *local* intra-pair spin permutations $(AA')$, $(MM')$, or $(XX')$, so that $T_i^{AA'}T_j^{MM'}T_k^{XX'}$ belongs to the local (*ggg*) manifold, whereas $S_0^{AA'}S_0^{MM'}T_i^{XX'}$ belongs to (*uug*).

In general, the Hamiltonian of an $AA'MM'XX'$ system can be split into terms having different local symmetries:

$$\hat{H} = \hat{H}_Z^{ggg} + \hat{H}_J^{ggg} + \hat{H}_J^{uug} + \hat{H}_J^{guu}, \tag{1}$$

$$\hat{H}_Z^{ggg} = \nu_A(\hat{I}_z^A + \hat{I}_z^{A'}) + \nu_M(\hat{I}_z^M + \hat{I}_z^{M'}) + \nu_X(\hat{I}_z^X + \hat{I}_z^{X'}); \tag{2}$$

$$\hat{H}_J^{ggg} = J_{AA'}\hat{\boldsymbol{I}}^A \cdot \hat{\boldsymbol{I}}^{A'} + J_{MM'}\hat{\boldsymbol{I}}^M \cdot \hat{\boldsymbol{I}}^{M'} + J_{XX'}\hat{\boldsymbol{I}}^X \cdot \hat{\boldsymbol{I}}^{X'} + \tag{3}$$

$$\tfrac{1}{2}\Sigma J_{AM}(\hat{I}_z^A + \hat{I}_z^{A'})(\hat{I}_z^M + \hat{I}_z^{M'}) + \tfrac{1}{2}\Sigma J_{MX}(\hat{I}_z^M + \hat{I}_z^{M'})(\hat{I}_z^X + \hat{I}_z^{X'}) ;$$

$$\hat{H}_J^{uug} = \tfrac{1}{2}\Delta J_{AM}(\hat{I}_z^A - \hat{I}_z^{A'})(\hat{I}_z^M - \hat{I}_z^{M'}); \tag{4}$$

$$\hat{H}_J^{guu} = \tfrac{1}{2}\Delta J_{MX}(\hat{I}_z^M - \hat{I}_z^{M'})(\hat{I}_z^X - \hat{I}_z^{X'}). \tag{5}$$

Here $\Sigma J_{AM} = J_{AM} + J_{AM'}$, and $\Delta J_{AM} = J_{AM} - J_{AM'}$, etc. Note that $J_{A'M} = J_{AM'}$, and $J_{AM} = J_{A'M'}$, etc., which is confirmed by analysis of experimental spectra (see Table S4). All terms in Eq. (2) (3) are *gerade* with respect to all possible *local* intra-pair spin permutations $(AA')$, $(MM')$, or $(XX')$; those in Eq. (4) and (5) are *ungerade*, which allows one to create the LLS. Due to structural simi-larities, all intrapair J-couplings have similar values so that it is convenient to consider their aver-age $J_{intra} = \tfrac{1}{3}(J_{AA'} + J_{MM'} + J_{XX'})$. Intrapair *J*-couplings "protect" LLS against coherent interconversion between intra-pair singlet and triplet states, since these couplings are much larger than those that break the symmetry $J_{intra} \gg \Delta J = \frac{1}{2}(\Delta J_{AM} + \Delta J_{MX})$.

So far, applications of the spin-lock induced crossing (SLIC) method [5,40] have been limited to monochromatic irradiation at a *single* radio-frequency. Among other applications, SLIC has been used for heteronuclear systems of [13]C enriched diphenylacetylene [6,41] and [15]N enriched azobenzene [10,42], where the four *ortho* protons (ignoring the *meta* and *para* protons) and the two [15]N nuclei constitute an $AA'A''A'''XX'$ system. Somewhat counter-intuitively, the application of a monochromatic SLIC sequence at the common shift $\nu_A$ of the four *ortho* protons excites an LLS that involves the two [15]N nuclei. This constitutes evidence for the mixing of LLS along chains of coupled nuclei.

In this work, we introduce *polychromatic* SLIC (poly-SLIC). In $AA'MM'XX'$ systems, one may choose to apply one, two or three carrier frequencies with weak *RF* amplitudes, yielding 9 possible methods for magnetization-to-singlet conversion via double-quantum (DQ) or single-quantum (SQ) level anti-crossings (LACs), namely:

1-3    Single-frequency SLIC with single-quantum LAC by irradiation at only one of the three shifts $\nu_A$ or $\nu_M$ or $\nu_X$ (single SLIC with SQ LAC)

4, 5    Double-frequency SLIC with single-quantum LAC by simultaneous irradiation at the shifts $\nu_A$ and $\nu_M$, or, equivalently at the shifts $\nu_M$ and $\nu_X$ (double SLIC with SQ LAC of two neighbouring CH$_2$ groups)

6    Double-frequency SLIC with single-quantum LAC by simultaneous irradiation at the shifts $\nu_A$ and $\nu_X$ of the two terminal CH$_2$ moieties (double SLIC with SQ LAC of two remote CH$_2$ groups),

7, 8    Double-frequency SLIC with double quantum LAC by simultaneous irradiation at two shifts $\nu_A$ and $\nu_M$, or, equivalently, at the shifts $\nu_M$ and $\nu_X$ (double SLIC with DQ LAC of two neighbouring CH$_2$ groups)

9    Triple-frequency SLIC with DQ LAC by simultaneous irradiation at three shifts $\nu_A$, $\nu_M$, and $\nu_X$.

The efficiencies (quantum yields) of these 9 methods, along with a discussion of the SQ and DQ LACs, are given in the Supplemental Material. As shown in Figure 1, the magnetization is first converted into LLS by one of 9 possible methods. The resulting LLS decays slowly in the relaxation interval $\tau_{rel}$. A $T_{00}$ filter [27] then eliminates all off-diagonal elements of the density matrix, except for zero-quantum coherences. The remaining terms are then re-converted back into observable magnetization, again by one of 9 possible methods (see Tables S2 and S3). The most efficient scheme uses triple SLIC excitation and single SLIC reconversion to maximize the observable magnetization of one of the spin pairs.

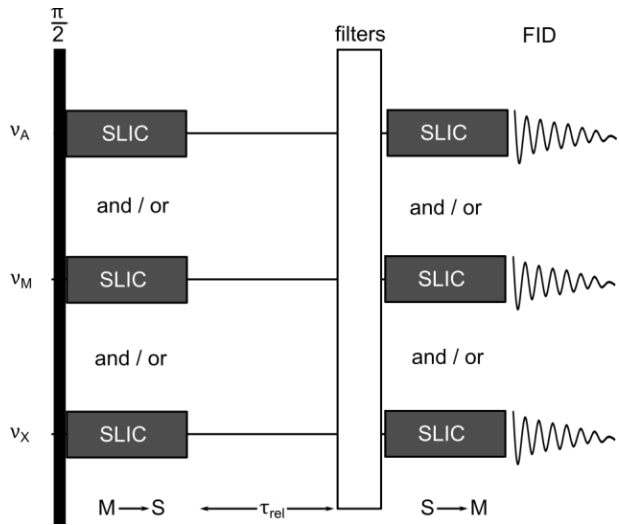

Figure 1. Pulse sequences for poly-SLIC applied to a $AA'MM'XX'$ system of a chain of 3 neighboring CH₂ groups. All 9 methods start by exciting transverse magnetization by a 'hard' $(\pi/2)_x$ pulse, followed by the simultaneous application of one, two or three selective RF fields applied simultaneously at resonance frequencies (chemical shifts) $\nu_A$ and/or $\nu_M$ and/or $\nu_X$ with phases $\pm y$. The poly-SLIC pulses convert the magnetization into LLS. The RF amplitudes must be $\nu_{RF} = 2J_{intra}$ to achieve LLS excitation by single-quantum level anti-crossing (SQ LAC), or $\nu_{RF} = J_{intra}$ for double-quantum (DQ) LAC. Maximum efficiency is achieved when $\tau_{SLIC}^{SQ} = 1/(|\sqrt{2}\Delta J|)$ for SQ LAC and $\tau_{SLIC}^{DQ} = \sqrt{2}\tau_{SLIC}^{SQ}$ for DQ LAC. After a $T_{00}$ filter [43], poly-SLIC sequences allow one to reconvert LLS into observable magnetization.

The Hamiltonian for one, two or three RF fields that are on-resonance with 2, 4 or 6 spins $p \in \{A, A', M, M', X, X'\}$ is given by:

$$\hat{H}_{RF}^{p} = -\nu_{RF}\left(\sum_p \hat{I}_y^p\right), \tag{6}$$

here $\nu_{RF}$ denotes the RF amplitude (Rabi frequency) expressed in Hz, the summation being taken only over the irradiated spins. The RF Hamiltonians applied to the different pairs commute with each other, so that the Zeeman terms in Eq. (2) can be neglected (in a triply rotating frame). Hence only $\hat{H}_J$ and $\hat{H}_{RF}$ in Eq. (6) contribute to the excitation of LLS. The level anti-crossings (LACs) shown in Figure S2 allow one to swap populations between states with different symmetries, provided the durations of the SLIC pulses are close to their optimum values, i.e., $\tau_{SLIC}^{SQ} = 1/(\left|\sqrt{2}\Delta J\right|)$ for SQ LAC and $\tau_{SLIC}^{DQ} = 1/(\left|\Delta J\right|)$ for DQ LAC.

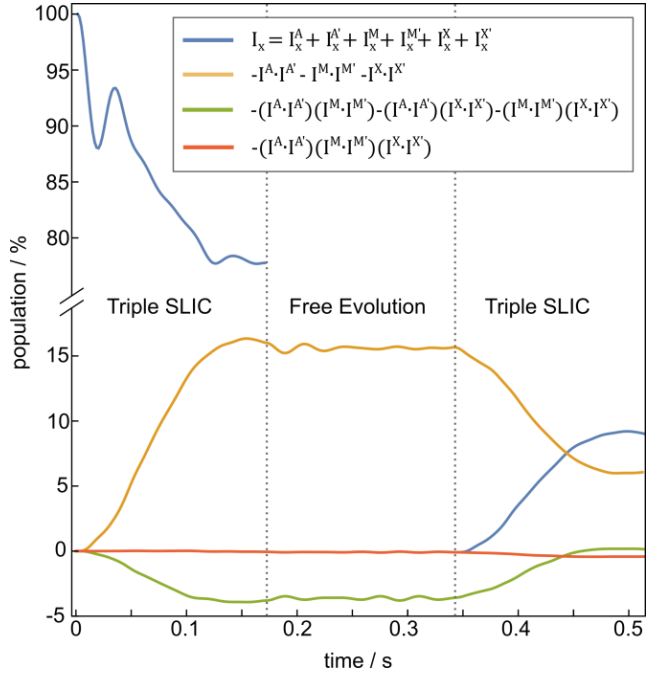

Figure 2. Numerical simulations of the excitation and reconversion of three types of delocalized long-lived states (LLS) in Eq. (7). The superposition of two-spin product terms (yellow) has coefficients above 15% after triple SLIC excitation, the four-spin product terms (green) have coefficients below 5%, while the six-spin product term (red) has a negligible amplitude. The blue lines correspond to the trajectory of the magnetization that is partly converted into LLS and back. All simulations were carried out with Spin Dynamica [44].

Numerical simulations have been performed to determine which LLS terms are excited by poly-SLIC (see Supplement). *Inter alia*, the following *delocalised* density operator terms are obtained:

$$\hat{\sigma}_{LLS} = -\lambda_{AA'}N_2\hat{I}^A \cdot \hat{I}^{A'} - \lambda_{MM'}N_2\hat{I}^M \cdot \hat{I}^{M'} - \lambda_{XX'}N_2\hat{I}^X \cdot \hat{I}^{X'} \tag{7}$$
$$- \lambda_{AA'MM'}N_4(\hat{I}^A \cdot \hat{I}^{A'})(\hat{I}^M \cdot \hat{I}^{M'}) - \lambda_{AA'XX'}N_4(\hat{I}^A \cdot \hat{I}^{A'})(\hat{I}^X \cdot \hat{I}^{X'})$$
$$- \lambda_{MM'XX'}N_4(\hat{I}^M \cdot \hat{I}^{M'})(\hat{I}^X \cdot \hat{I}^{X'}) - \lambda_{AA'MM'XX'}N_6(\hat{I}^A \cdot \hat{I}^{A'})(\hat{I}^M \cdot \hat{I}^{M'})(\hat{I}^X \cdot \hat{I}^{X'}),$$

with $N_2 = \frac{1}{2\sqrt{3}}$, $N_4 = \frac{2}{3}$, $N_6 = \frac{8}{3\sqrt{3}}$. The bilinear terms are familiar products of $m = 2$ spin operators, while the higher terms contain unusual products of $m = 4$ and 6 spin operators. All operators in Eq. (7) correspond to population imbalances. Within the *gerade* manifold, $\hat{\sigma}_{LLS}$ comprises an imbalance between the mean population of the 9 states (*uug, ugu, guu*) and the mean population of the 27 states (*ggg*). Within the *ungerade* manifold, there is an imbalance between the population of the unique singlet (*uuu*) state and the mean population of the 27 states (*ugg, gug, ggu*). Although many other terms are excited, they relax rapidly or can be eliminated by filtration and phase cycling. The commutator between the average two spin-terms $1/3 \left(\lambda_{AA'} + \lambda_{MM'} + \right.$

$\lambda_{XX'})N_2\left(\hat{I}^A \cdot \hat{I}^{A'} + \hat{I}^M \cdot \hat{I}^{M'} + \hat{I}^X \cdot \hat{I}^{X'}\right)$ and the free-precession Hamiltonian yields a negligible residue. The same is true for the sum of the three four-spin terms, as well as for the six-spin term. Therefore, these terms are nearly invariant under free evolution, as evidenced by the weak oscillations during free evolution shown in Figure 2.

One should distinguish two aspects of the delocalized nature of the LLS: (i) delocalization by mixing of two- and four-spin states, and (ii) delocalization by excitation of products comprising four- and six-spin states. A manifestation of *delocalization by mixing* is the experimental fact that a monochromatic SLIC pulse applied only to the $AA'$ pair also excites LLS associated with the $MM'$ and $XX'$ pairs (Figure 4). The second (more remarkable) effect is that we can create LLS that are delocalized over *n* neighboring $CH_2$ groups because the density operator is made up of a *product*, rather than a sum, of population imbalances, such as the six-spin term $\left(\hat{I}^A \cdot \hat{I}^{A'}\right)\left(\hat{I}^M \cdot \hat{I}^{M'}\right)\left(\hat{I}^X \cdot \hat{I}^{X'}\right)$. Such states are rarely observed in NMR and may be of interest for quantum information processing.

Figure 3 shows a selection of experimental multiplets typical of the $AA'MM'XX'$ system of $NaSO_3CH_2CH_2CH_2Si(CH_3)_3$ (2,2-Dimethyl-2-silapentane-5-sulfonate sodium salt, also known as DSS), the methyl groups of which are commonly used as a chemical shift standard in aqueous solution. The multiplets obtained after conversion of magnetization into LLS and back are characteristic of the terms contained in the density operator. Their integrated intensities reach *ca.* 6% of a conventional spectrum. The agreement between simulated and experimental multiplets is good.

A dramatic simplification of multiplets such as those shown in Figure 3 can be achieved by polychromatic homonuclear decoupling during the observation of the FID's [45], so that each multiplet collapses to a single line, as will be shown elsewhere.

Polychromatic Excitation of Delocalized Long-Lived Proton Spin States in Aliphatic Chains

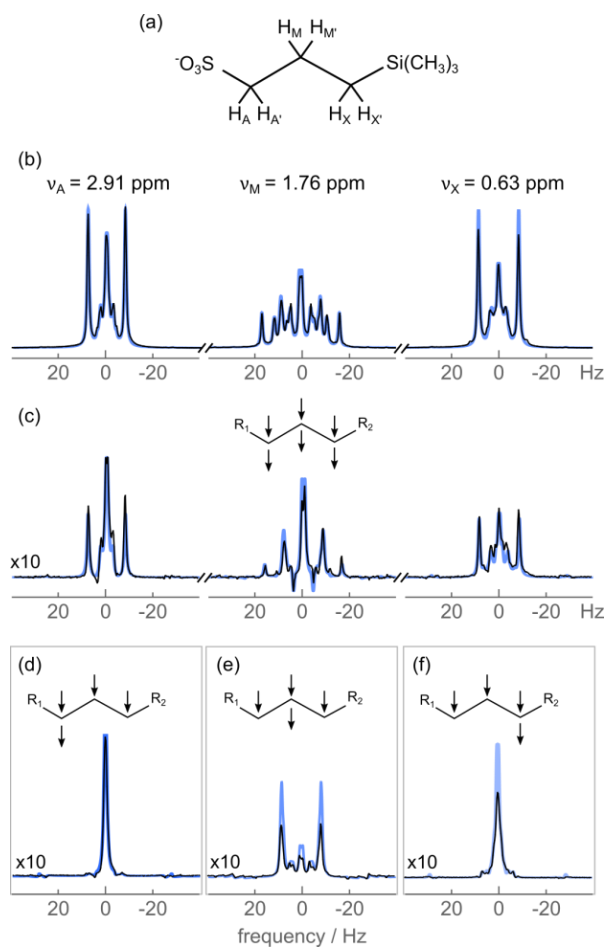

Figure 3. (a) Molecular structure of $NaSO_3CH_2CH_2CH_2Si(CH_3)_3$ (DSS). (b) Multiplets of its $AA'MM'XX'$ proton system, excited by a 'hard' $\pi/2$ pulse applied to thermal equilibrium (black: experimental multiplets, blue: simulated multiplets of the strongly coupled 6 spin system [46]) (c) Multiplets obtained by triple SLIC excitation of LLS followed, after a relaxation interval $\tau_{rel}$ = 3 s and a $T_{00}$ filter, by reconversion into observable magnetization by triple SLIC irradiation. The arrows indicate at which chemical shifts the SLIC irradiation is applied. (d-f) Multiplets obtained in three separate experiments, each after triple SLIC excitation, evolution during $\tau_{rel}$ = 3 s, and a $T_{00}$ filter. (d) Multiplet of the $AA'$ pair obtained by single SLIC reconversion by irradiation at $\nu_A$; (e) multiplet of the $MM'$ pair after irradiation at $\nu_M$; (f) multiplet of the $XX'$ pair after irradiation at $\nu_X$. Note the agreement between experimental and simulated spectra (black and blue lines respectively). The vertical scales of the experimental multiplets in (c)-(f) were increased by factor 10. All the spectra were obtained by addition of 8 transients.

Figure 4 shows decays of LLS as a function of the relaxation interval $\tau_{rel}$. Four experiments have been performed with different SLIC excitation pulses and with the same reconversion SLIC applied at the $\nu_A$ frequency. The relaxation rates of LLS are dominated by out-of-pair dipole-dipole couplings, so that these rates are reduced by the 6th power ratio of the internuclear distances $(r_{AM}/r_{AA'})^6$ with respect to the longitudinal rates $1/T_1$ that are mostly determined by intra-pair dipole-dipole couplings. Triple SLIC excitation gives amplitudes that are almost twice as large as single SLIC. The amplitude of LLS signal excited by SLIC applied at the $\nu_X$ frequency gives the smallest amplitude, but it illustrates the delocalized character of created LLS, which results from the mixing of spin states. Experiment where both the excitation and the reconversion SLIC pulses are applied at one and the same frequency $\nu_A$ (green diamonds in Figure 4) gives a $T_{LLS}$ that is slightly shorter than the other experiments which yield similar $T_{LLS}$. This may be attributed to the fact that in this case the LLS is predominantly localized on the AA' pair (see tables S1, S2).

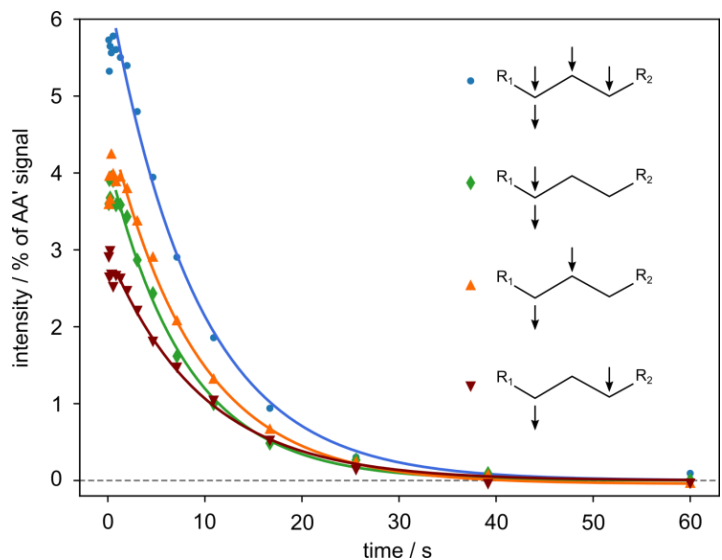

Figure 4. Typical decays of long-lived states. The arrows indicate at which sites the SLIC irradiation is applied. (Blue dots) LLS excitation by triple SLIC, reconversion by single SLIC at $\nu_A$ only ($T_{LLS} = 9.3 \pm 0.4$ s). (Green diamonds) Excitation and reconversion both by single SLIC at $\nu_A$ ($T_{LLS} = 8.0 \pm 0.4$ s). (Orange triangles) Excitation by single SLIC at $\nu_M$ and reconversion on neighboring group by single SLIC at $\nu_A$ ($T_{LLS} = 9.6 \pm 0.5$ s). (Red downward pointing triangles) Excitation by single SLIC at $\nu_X$ and remote reconversion by single SLIC at $\nu_A$ ($T_{LLS} = 9.7 \pm 0.4$ s). The solid lines correspond to mono-exponential fits for $\tau_{rel} > 0.84$ s, ignoring initial oscillations. The largest amplitude is observed for triple SLIC excitation. The LLS lifetimes are at least 5 times longer than the corresponding longitudinal relaxation times $T_1^A \approx T_1^M \approx T_1^X \approx 1.5 \pm 0.03$ s.

It is straightforward to excite LLS in potential drug molecules that contain neighboring $CH_2$ groups. The contrast between the relaxation rates $R_{LLS} = 1/T_{LLS}$ of free and partly bound drug molecules allows one to determine the affinity (binding constants) for macromolecules such as proteins [16–18]. Binding may partly lift the chemical equivalence, i.e., convert an $AA'MM'XX'$ system into an $ABMNXY$ system. This will contribute to a leakage between intra-pair $T_0$ and $S_0$ states, thus levelling out the population imbalances, reducing $T_{LLS}$, and enhancing the contrast between free and bound molecules. By way of example, 5 mM DSS has been titrated with bovine serum albumin (BSA) over the range 0.5 μM ≤ [BSA] ≤ 20 μM. As shown in the Supplemental Material, the lifetime of the LLS of DSS drops steeply by almost factor of 10 upon addition of BSA at very low concentrations [BSA]/[DSS] > $10^{-4}$, whereas the $^1H$ chemical shifts, $T_1$ and $T_2$ are barely affected for [BSA] < 20 µM. The contrast could be amplified by shuttling the sample to a lower field during the relaxation interval $\tau_{rel}$ [47].

The novel poly-SLIC method applied to aliphatic chains opens several prospects for future work. In a chain with $n$ $CH_2$ groups, various delocalised states can be addressed via different entrance ports, and can be read out at the same or other output ports. Similar LLS can be found in many chemical compounds, e.g., in neurotransmitters like gamma-aminobutyric acid (GABA).

To conclude, delocalised long-lived states encompassing all 6 proton spins in $AA'MM'XX'$ systems in aliphatic chains of neighbouring $CH_2$ groups have been excited and observed by poly-chromatic spin-lock induced crossing (poly-SLIC). The lifetimes $T_{LLS}$ of the long-lived states are typically 5 times longer than $T_1$. The lifetimes $T_{LLS}$ are significantly shortened when molecules that carry long-lived states interact with macromolecules such as proteins. This contrast should be useful for screening of drug molecules without the need for labelling with fluorescent tags or other chemical modifications.

### Acknowledgements

We are indebted to Dr Philippe Pelupessy for stimulating discussions, to the CNRS and the ENS for support, and to the European Research Council (ERC) for the Synergy grant "Highly Informative Drug Screening by Overcoming NMR Restrictions" (HISCORE, grant agreement number 951459).

### References

[1] M. Carravetta, O. G. Johannessen, and M. H. Levitt, *Beyond the T1 Limit: Singlet Nuclear Spin States in Low Magnetic Fields*, Phys. Rev. Lett. **92**, 153003 (2004).

[2] M. Carravetta and M. H. Levitt, *Long-Lived Nuclear Spin States in High-Field Solution NMR*, J. Am. Chem. Soc. **126**, 6228 (2004).

[3] R. Sarkar, P. R. Vasos, and G. Bodenhausen, *Singlet-State Exchange NMR Spectroscopy for the Study of Very Slow Dynamic Processes*, J. Am. Chem. Soc. **129**, 328 (2007).

[4] M. C. D. Tayler and M. H. Levitt, *Singlet Nuclear Magnetic Resonance of Nearly-Equivalent Spins*, Phys. Chem. Chem. Phys. **13**, 5556 (2011).

[5] S. J. DeVience, R. L. Walsworth, and M. S. Rosen, *Preparation of Nuclear Spin Singlet States Using Spin-Lock Induced Crossing*, Phys. Rev. Lett. **111**, 173002 (2013).

[6] Y. Feng, T. Theis, X. Liang, Q. Wang, P. Zhou, and W. S. Warren, *Storage of Hydrogen Spin Polarization in Long-Lived 13C2 Singlet Order and Implications for Hyperpolarized Magnetic Resonance Imaging*, J. Am. Chem. Soc. **135**, 9632 (2013).

[7] K. Claytor, T. Theis, Y. Feng, J. Yu, D. Gooden, and W. S. Warren, *Accessing Long-Lived Disconnected Spin-1/2 Eigenstates through Spins > 1/2*, J. Am. Chem. Soc. **136**, 15118 (2014).

[8] G. Stevanato, S. S. Roy, J. Hill-Cousins, I. Kuprov, L. J. Brown, R. C. D. Brown, G. Pileio, and M. H. Levitt, *Long-Lived Nuclear Spin States Far from Magnetic Equivalence*, Phys. Chem. Chem. Phys. **17**, 5913 (2015).

[9] A. N. Pravdivtsev, A. S. Kiryutin, A. V. Yurkovskaya, H.-M. Vieth, and K. L. Ivanov, *Robust Conversion of Singlet Spin Order in Coupled Spin-1/2 Pairs by Adiabatically Ramped RF-Fields*, J. Magn. Reson. **273**, 56 (2016).

[10] K. F. Sheberstov, H.-M. Vieth, H. Zimmermann, B. A. Rodin, K. L. Ivanov, A. S. Kiryutin, and A. V. Yurkovskaya, *Generating and Sustaining Long-Lived Spin States in 15 N, 15 N'-Azobenzene*, Sci. Rep. **9**, 1 (2019).

[11] P. R. Vasos, A. Comment, R. Sarkar, P. Ahuja, S. Jannin, J.-P. Ansermet, J. A. Konter, P. Hautle, B. van den Brandt, and G. Bodenhausen, *Long-Lived States to Sustain Hyperpolarized Magnetization*, PNAS **106**, 18469 (2009).

[12] D. B. Burueva, J. Eills, J. W. Blanchard, A. Garcon, R. Picazo-Frutos, K. V. Kovtunov, I. V. Koptyug, and D. Budker, *Chemical Reaction Monitoring Using Zero-Field Nuclear Magnetic Resonance Enables Study of Heterogeneous Samples in Metal Containers*, Angew. Chem. Int. Ed. Engl. **59**, 17026 (2020).

[13] C. Bengs, L. Dagys, G. A. I. Moustafa, J. W. Whipham, M. Sabba, A. S. Kiryutin, K. L. Ivanov, and M. H. Levitt, *Nuclear Singlet Relaxation by Chemical Exchange*, J. Chem. Phys. **155**, 124311 (2021).

[14] S. Cavadini, J. Dittmer, S. Antonijevic, and G. Bodenhausen, *Slow Diffusion by Singlet State NMR Spectroscopy*, J. Am. Chem. Soc. **127**, 15744 (2005).

[15] M. C. Tourell, I.-A. Pop, L. J. Brown, R. C. D. Brown, and G. Pileio, *Singlet-Assisted Diffusion-NMR (SAD-NMR): Redefining the Limits When Measuring Tortuosity in Porous Media*, Phys. Chem. Chem. Phys. **20**, 13705 (2018).

[16] N. Salvi, R. Buratto, A. Bornet, S. Ulzega, I. Rentero Rebollo, A. Angelini, C. Heinis, and G. Bodenhausen, *Boosting the Sensitivity of Ligand–Protein Screening by NMR of Long-Lived States*, J. Am. Chem. Soc. **134**, 11076 (2012).

[17] R. Buratto, D. Mammoli, E. Chiarparin, G. Williams, and G. Bodenhausen, *Exploring Weak Ligand– Protein Interactions by Long-Lived NMR States: Improved Contrast in Fragment-Based Drug Screening*, Angew. Chem. Int. Ed. **53**, 11376 (2014).

[18] R. Buratto, D. Mammoli, E. Canet, and G. Bodenhausen, *Ligand–Protein Affinity Studies Using Long-Lived States of Fluorine-19 Nuclei*, J. Med. Chem. **59**, 1960 (2016).

[19] G. Pileio, M. Carravetta, and M. H. Levitt, *Extremely Low-Frequency Spectroscopy in Low-Field Nuclear Magnetic Resonance*, Phys. Rev. Lett. **103**, 083002 (2009).

[20] R. Sarkar, P. Ahuja, P. R. Vasos, and G. Bodenhausen, *Long-Lived Coherences for Homogeneous Line Narrowing in Spectroscopy*, Phys. Rev. Lett. **104**, 053001 (2010).

[21] K. F. Sheberstov, A. S. Kiryutin, C. Bengs, J. T. Hill-Cousins, L. J. Brown, R. C. D. Brown, G. Pileio, M. H. Levitt, A. V. Yurkovskaya, and K. L. Ivanov, *Excitation of Singlet–Triplet Coherences in Pairs of Nearly-Equivalent Spins*, Phys. Chem. Chem. Phys. **21**, 6087 (2019).

[22] B. A. Rodin, C. Bengs, A. S. Kiryutin, K. F. Sheberstov, L. J. Brown, R. C. D. Brown, A. V. Yurkovskaya, K. L. Ivanov, and M. H. Levitt, *Algorithmic Cooling of Nuclear Spins Using Long-Lived Singlet Order*, J. Chem. Phys. **152**, 164201 (2020).

[23] G. Pileio, *Relaxation Theory of Nuclear Singlet States in Two Spin-1/2 Systems*, Prog. Nucl. Magn. Reson. Spectrosc. **56**, 217 (2010).

[24] G. Pileio and M. H. Levitt, *J-Stabilization of Singlet States in the Solution NMR of Multiple-Spin Systems*, J. Magn. Reson. **187**, 141 (2007).

[25] M. B. Franzoni, L. Buljubasich, H. W. Spiess, and K. Münnemann, *Long-Lived 1H Singlet Spin States Originating from Para-Hydrogen in Cs-Symmetric Molecules Stored for Minutes in High Magnetic Fields*, J. Am. Chem. Soc. **134**, 10393 (2012).

[26] M. C. D. Tayler, I. Marco-Rius, M. I. Kettunen, K. M. Brindle, M. H. Levitt, and G. Pileio, *Direct Enhancement of Nuclear Singlet Order by Dynamic Nuclear Polarization*, J. Am. Chem. Soc. **134**, 7668 (2012).

[27] D. Mammoli et al., *Challenges in Preparing, Preserving and Detecting Para -Water in Bulk: Overcoming Proton Exchange and Other Hurdles*, Phys. Chem. Chem. Phys. **17**, 26819 (2015).

[28] G. Pileio, M. Concistrè, M. Carravetta, and M. H. Levitt, *Long-Lived Nuclear Spin States in the Solution NMR of Four-Spin Systems*, J. Magn. Reson. **182**, 353 (2006).

[29] E. Vinogradov and A. K. Grant, *Hyperpolarized Long-Lived States in Solution NMR: Three-Spin Case Study in Low Field*, J. Magn. Reson. **194**, 46 (2008).

[30] A. K. Grant and E. Vinogradov, *Long-Lived States in Solution NMR: Theoretical Examples in Three- and Four-Spin Systems*, J. Magn. Reson. **193**, 177 (2008).

[31] P. Ahuja, R. Sarkar, P. R. Vasos, and G. Bodenhausen, *Long-Lived States in Multiple-Spin Systems*, ChemPhysChem **10**, 2217 (2009).

[32] H. J. Hogben, P. J. Hore, and I. Kuprov, *Multiple Decoherence-Free States in Multi-Spin Systems*, J. Magn. Reson. **211**, 217 (2011).

[33] G. Pileio, J. T. Hill-Cousins, S. Mitchell, I. Kuprov, L. J. Brown, R. C. D. Brown, and M. H. Levitt, *Long-Lived Nuclear Singlet Order in Near-Equivalent 13C Spin Pairs*, J. Am. Chem. Soc. **134**, 17494 (2012).

[34] G. Stevanato, J. T. Hill-Cousins, P. Håkansson, S. S. Roy, L. J. Brown, R. C. D. Brown, G. Pileio, and M. H. Levitt, *A Nuclear Singlet Lifetime of More than One Hour in Room-Temperature Solution*, Angew. Chem. Int. Ed. Engl. **54**, 3740 (2015).

[35] S. J. Elliott, P. Kadeřávek, L. J. Brown, M. Sabba, S. Glöggler, D. J. O'Leary, R. C. D. Brown, F. Ferrage, and M. H. Levitt, *Field-Cycling Long-Lived-State NMR of 15N2 Spin Pairs*, Mol. Phys. **117**, 861 (2019).

[36] K. Shen, A. W. J. Logan, J. F. P. Colell, J. Bae, G. X. Ortiz, T. Theis, W. S. Warren, S. J. Malcolmson, and Q. Wang, *Diazirines as Potential Molecular Imaging Tags: Probing the Requirements for Efficient and Long-Lived SABRE-Induced Hyperpolarization*, Angew. Chem. Int. Ed. Engl. **56**, 12112 (2017).

[37] J. Eills, G. Stevanato, C. Bengs, S. Glöggler, S. J. Elliott, J. Alonso-Valdesueiro, G. Pileio, and M. H. Levitt, *Singlet Order Conversion and Parahydrogen-Induced Hyperpolarization of 13C Nuclei in near-Equivalent Spin Systems*, J. Magn. Reson. **274**, 163 (2017).

[38] G. Schrumpf, *NMR Spectra of Monosubstituted Alkanes I. N-Propyl Derivatives*, J. Magn. Reson. (1969) **6**, 243 (1972).

[39] J. A. Pople, W. G. Schneider, and H. J. Bernstein, *The Analysis of Nuclear Magnetic Resonance Spectra: Ii. Two Pairs of Two Equivalent Nuclei*, Can. J. Chem. **35**, 1060 (1957).

[40] S. J. DeVience, M. Greer, S. Mandal, and M. S. Rosen, *Homonuclear J-Coupling Spectroscopy at Low Magnetic Fields Using Spin-Lock Induced Crossing\*\**, ChemPhysChem **22**, 2128 (2021).

[41] Y. Feng, T. Theis, T.-L. Wu, K. Claytor, and W. S. Warren, *Long-Lived Polarization Protected by Symmetry*, J. Chem. Phys. **141**, 134307 (2014).

[42] K. F. Sheberstov, H.-M. Vieth, H. Zimmermann, K. L. Ivanov, A. S. Kiryutin, and A. V. Yurkovskaya, *Cis Versus Trans-Azobenzene: Precise Determination of NMR Parameters and Analysis of Long-Lived States of 15N Spin Pairs*, Appl. Magn. Reson. **49**, 293 (2018).

[43] M. C. D. Tayler, *Chapter 10: Filters for Long-Lived Spin Order*, in *Long-Lived Nuclear Spin Order* (The Royal Society of Chemistry, 2020), pp. 188–208.

[44] C. Bengs and M. H. Levitt, *SpinDynamica: Symbolic and Numerical Magnetic Resonance in a Mathematica Environment*, Magn. Reson. Chem. **56**, 374 (2018).

[45] D. Carnevale, T. F. Segawa, and G. Bodenhausen, *Polychromatic Decoupling of a Manifold of Homonuclear Scalar Interactions in Solution-State NMR*, Eur. J. Chem. **18**, 11573 (2012).

[46] D. A. Cheshkov, K. F. Sheberstov, D. O. Sinitsyn, and V. A. Chertkov, *ANATOLIA: NMR software for spectral analysis of total lineshape*, Magn. Reson. Chem. **56**, 449 (2018).

[47] Z. Wang et al., *Detection of Metabolite–Protein Interactions in Complex Biological Samples by High-Resolution Relaxometry: Toward Interactomics by NMR*, J. Am. Chem. Soc. **143**, 9393 (2021).

# Supplemental Material

**Polychromatic Excitation of Delocalized Long-Lived Proton Spin States in Aliphatic Chains**

Anna Sonnefeld, Geoffrey Bodenhausen, Kirill Sheberstov*

*Laboratoire des biomolécules, LBM, Département de chimie, École normale supérieure, PSL University, Sorbonne Université, CNRS, 75005 Paris, France*

*Corresponding author: kirill.sheberstov@ens.psl.eu

**Contents**

### Hamiltonian of the AA'MM'XX' system

Figure S1 shows a matrix representation of the Hamiltonian $\hat{H}_J = \hat{H}_J^{ggg} + \hat{H}_J^{uug} + \hat{H}_J^{guu}$ (Eqs. 3 - 4 of the main text) in the intra-pair singlet-triplet product basis. In the upper left block, there are 6 off-diagonal elements with values of either $\frac{1}{2}\Delta J_{AM}$ or $\frac{1}{2}\Delta J_{MX}$ on each side of the diagonal between states with different local symmetry, plus another 2 off-diagonal elements in the lower right block (all 16 elements are highlighted by red squares Figure S1). These off-diagonal elements provide access LLS by SLIC. The differences between the corresponding diagonal elements are of the order of twice the intra-pair $2J_{intra}$ couplings, i.e., much larger than the off-diagonal elements that are of the order of $\frac{1}{2}\Delta J_{AM} \approx \frac{1}{2}\Delta J_{MX}$. Therefore, the product states are good approximations for the true eigenstates.

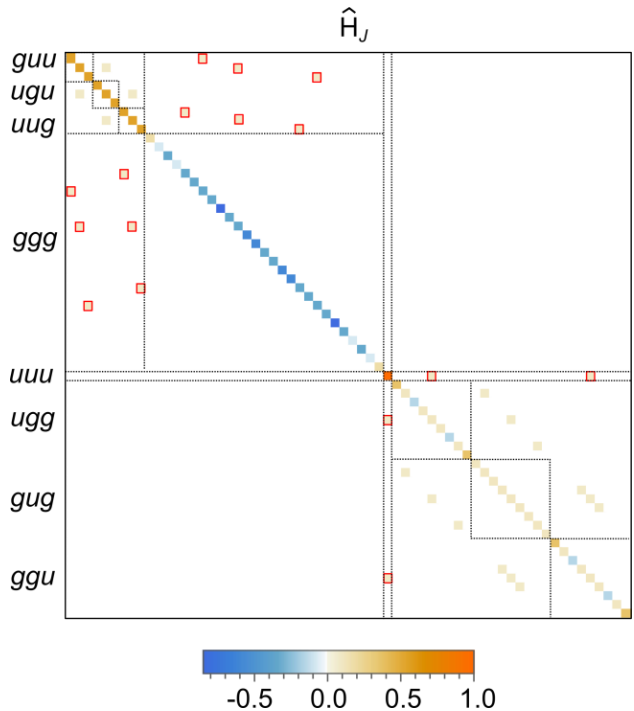

Figure S1. Matrix representation of the Hamiltonian $\hat{H}_J$ of the $AA'MM'XX'$ system in a basis obtained by direct products of intra-pair singlet and triplet states of the $AA'$, $MM'$ and $XX'$ pairs. The states are sorted according to their ($u$) or ($g$) symmetry under local intra-pair spin permutations. The 16 off-diagonal elements highlighted by red squares connect states that belong to different local ($u$) and ($g$) symmetries and allow one to access LLS by SLIC. The color scale gives the size of the matrix elements $-0.8 \leq H_{ij} \leq +1$, normalized with respect to the unique ($uuu$) state $S_0^{AA'}\,S_0^{MM'}\,S_0^{XX'}$ which has the largest value. All off-diagonal elements within the upper left (globally *gerade*) manifold and the lower right (globally *ungerade*) blocks that are not highlighted can be diagonalized analytically.

Of the 36 states within the upper left globally *gerade* block, 11 are connected by 6 off-diagonal elements:

$$\left\{\begin{array}{l} T_{+1}^{AA'} S_0^{MM'} S_0^{XX'} \leftrightarrow T_{+1}^{AA'} T_0^{MM'} T_0^{XX'}, \\ S_0^{AA'} S_0^{MM'} T_{+1}^{XX'} \leftrightarrow T_0^{AA'} T_0^{MM'} T_{+1}^{XX'} \end{array}\right\},$$

$$(S1)$$

$$\{T_0^{AA'}S_0^{MM'}S_0^{XX'} \leftrightarrow T_{+1}^{AA'}T_0^{MM'}T_{+1}^{XX'} \leftrightarrow S_0^{AA'}S_0^{MM'}T_0^{XX'}\},$$

$$\left.\begin{array}{l} T_{-1}^{AA'}S_0^{MM'}S_0^{XX'} \leftrightarrow T_{-1}^{AA'}T_0^{MM'}T_0^{XX'}, \\ S_0^{AA'}S_0^{MM'}T_{-1}^{XX'} \leftrightarrow T_0^{AA'}T_0^{MM'}T_{-1}^{XX'} \end{array}\right\},$$

where double-headed arrows represent off-diagonal elements of the Hamiltonian $\hat{H}_J^{uug}$ and $\hat{H}_J^{guu}$ (see Eqs. 4 and 5 in the main text). Of the 28 states within the globally *ungerade* lower manifold, only 3 are connected by 2 off-diagonal elements:

$$\{T_0^{AA'}T_0^{MM'}S_0^{XX'} \leftrightarrow S_0^{AA'}S_0^{MM'}S_0^{XX'} \leftrightarrow T_0^{AA'}T_0^{MM'}S_0^{XX'}\}. \tag{S2}$$

None of the 16 other off-diagonal elements in the upper left 9x9 block and the lower 27x27 block of $\hat{H}_J$ can be neglected. The diagonalization of these blocks in the Hamiltonian can be achieved by simple unitary transformations. There are two 3x3 blocks, one in the globally *gerade* manifold:

$$\{T_0^{AA'}S_0^{MM'}S_0^{XX'} \leftrightarrow S_0^{AA'}T_0^{MM'}S_0^{XX'} \leftrightarrow S_0^{AA'}S_0^{MM'}T_0^{XX'}\} \tag{S3}$$

And the other in the globally *ungerade* manifold:

$$\{T_0^{AA'}T_0^{MM'}S_0^{XX'} \leftrightarrow T_0^{AA'}S_0^{MM'}T_0^{XX'} \leftrightarrow T_0^{AA'}T_0^{MM'}S_0^{XX'}\} \tag{S4}$$

Both can be diagonalized by the unitary transformation:

$$U_{3x3} = \frac{1}{2}\begin{pmatrix} 1 & \sqrt{2} & 1 \\ \sqrt{2} & 0 & \sqrt{2} \\ 1 & -\sqrt{2} & -1 \end{pmatrix}. \tag{S5}$$

Finally, there are four 2x2 blocks between globally *ungerade* states:

$$\{S_0^{AA'}T_0^{MM'}T_{+1}^{XX'} \leftrightarrow T_0^{AA'}S_0^{MM'}T_{+1}^{XX'}\},$$
$$\{S_0^{AA'}T_0^{MM'}T_{-1}^{XX'} \leftrightarrow T_0^{AA'}S_0^{MM'}T_{-1}^{XX'}\},$$
$$\{T_{+1}^{AA'}T_0^{MM'}S_0^{XX'} \leftrightarrow T_{+1}^{AA'}S_0^{MM'}T_0^{XX'}\},$$
$$\{T_{-1}^{AA'}T_0^{MM'}S_0^{XX'} \leftrightarrow T_{-1}^{AA'}S_0^{MM'}T_0^{XX'}\}. \tag{S6}$$

All of them are diagonalized by a simple unitary transformation:

$$U_{2x2} = \frac{1}{\sqrt{2}}\begin{pmatrix} 1 & 1 \\ 1 & -1 \end{pmatrix}. \tag{S7}$$

This leads to mixing of the states of the intra-pair singlet-triplet product basis.

An important consequence is that the eigenfunctions of the Hamiltonian are *delocalized* along the chain. Thus, one of the delocalized eigenstates is:

$$\frac{1}{2}T_0^{AA'}T_0^{MM'}S_0^{XX'} + \frac{1}{\sqrt{2}}T_0^{AA'}S_0^{MM'}T_0^{XX'} + \frac{1}{2}T_0^{AA'}T_0^{MM'}S_0^{XX'} \tag{S8}$$

Eigenstates described by such linear combinations cannot be attributed to any particular spin pair, since they are delocalized across the entire 6-spin system. Consequently, when an LLS is created by irradiating spin pair $AA'$ there is always some LLS that is associated with the other pairs $MM'$ and $XX'$.

Note that the resulting basis is not the exact eigenbasis, since the 16 off-diagonal elements highlighted by red squares in Figure S1 have been neglected, which is justified by perturbation theory, since $2J_{intra} >> \frac{1}{2}\Delta J_{AM} \approx \frac{1}{2}\Delta J_{MX}$.

### Level anti-crossings under polychromatic SLIC pulses

Figure S2 shows the energy levels of the 6-spin $AA'MM'XX'$ system during triple *RF* irradiation, calculated as a function of the *RF* field amplitude $\nu_{RF}$, which is the same for all three selective RF fields. The eigenstates of the Hamiltonian obtained by diagonalization in the presence of the *RF* fields are known as 'dressed states'. Since the intra-pair couplings in all three CH₂ groups are similar, we consider their average value $J_{intra} = \frac{1}{3}(J_{AA'} + J_{MM'} + J_{XX'})$. The key differences between the out-of-pair *J*-couplings that allow one to excite LLS are approximated by their average values $\Delta J = \frac{1}{2}(\Delta J_{AM} + \Delta J_{MX})$. The experimental couplings in Table S4 show that these assumptions are reasonable.

A dressed state is characterized by an effective total spin $I_\Sigma$ and its projections $m_\Sigma$. The four sets of lines in Figure S2 correspond to dressed states with effective total spin $I_\Sigma = 3$ and 2 (orange) and $I_\Sigma = 1$ or 0 (blue). For $I_\Sigma = 3$, there are 7 levels with $m_\Sigma = -3, -2, \dots, +3$. When $\nu_{RF}$ matches the conditions $\nu_{RF} = J_{intra}$, level anti-crossings (LACs) occur between dressed states with effective total spin $I_\Sigma = 3$ and $I_\Sigma = 1$ in the globally *gerade* manifold (Figure S2a). Another LAC is found in the globally *ungerade* block between the dressed states with effective total spin $I_\Sigma = 2$ and 0 (Figure S2b).

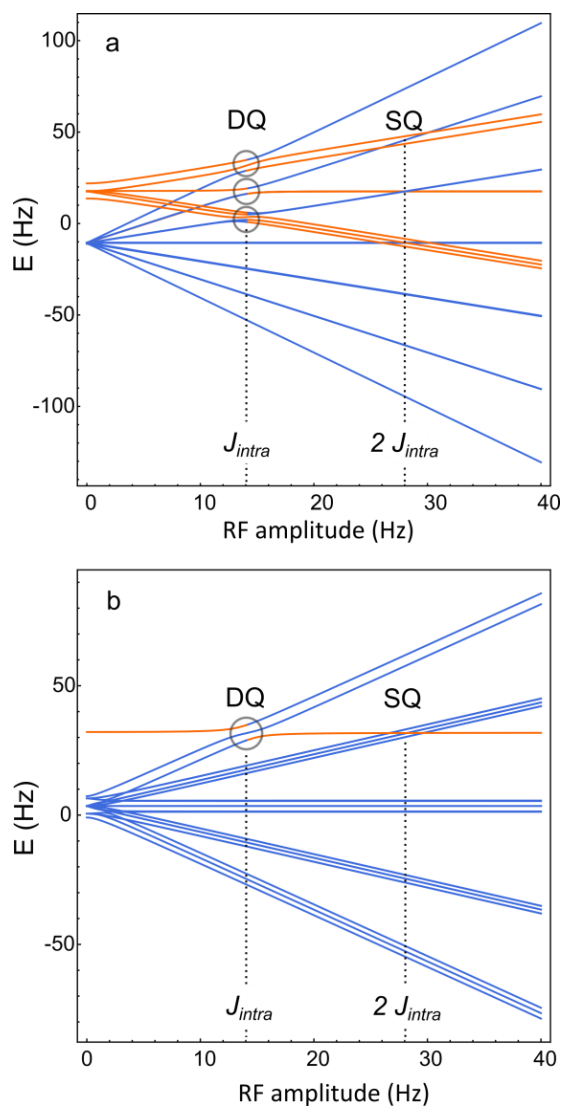

Figure S2. Energy levels of an $AA'MM'XX'$ system as a function of the *RF* amplitude $\nu_{RF}$ that is common to the three selective *RF* fields under triple SLIC irradiation. Level anti-crossings (LACs) where singlet and triplet populations can be swapped after suitable durations of *RF* irradiation are highlighted by circles. (a) Energy levels belonging to the globally *gerade* manifold of spin functions (e.g., $ggg$, $uug$, $ugu$, $guu$). The set of 7 blue lines corresponds to dressed states of a fictious particle with total spin $I_\Sigma = 3$ comprising only triplet product states of symmetry $ggg$. The upper set of orange lines corresponds to $uug$, $ugu$, $guu$ states. (b) Energy levels within the globally *ungerade* manifold ($uuu$, $ugg$, $gug$, $ggu$). The set of blue lines corresponds to ($ugg$, $gug$, $ggu$) states, one of which has a LAC with the antisymmetric $uuu$ state represented by the orange line. Under triple SLIC irradiation only double-quantum (DQ) LACs occur. The optimum *RF* amplitude is $\nu_{RF} = J_{intra} \approx 14$ Hz. For the sake of clarity, some levels that intersect with the LAC regions are not shown, and it was assumed without loss of generality that $\Sigma J_{AM} = 0$ so that states with the same $m_\Sigma$ appear degenerate.

For triple SLIC, the LACs only occur between levels with $\Delta m_\Sigma = \pm 2$, which will henceforth be referred to as double-quantum (DQ) LACs. For double irradiation SLIC, single-quantum (SQ) LACs with $\Delta m_\Sigma = \pm 1$ occur in the vicinity of $\nu_{RF} = 2J_{intra}$, while double-quantum (DQ) LACs with $m_\Sigma = \pm 2$ occur near $\nu_{RF} = J_{intra}$. For single (monochromatic) SLIC, only single-quantum (SQ) LACs with $\Delta m_\Sigma = \pm 1$ occur when $\nu_{RF} = 2J_{intra}$. Note that, regardless of the number of selective RF fields, one cannot induce any triple-quantum (TQ) transitions in aliphatic chains, since the long-range couplings are negligible ($J_{AX} \approx 0$).

If the number of RF fields and their amplitudes are chosen to fulfill the conditions for SLIC with SQ LAC, a complete interconversion of the populations of states with different local *u* and *g* symmetries can be achieved by RF irradiation of duration $\tau_{SLIC}^{SQ} = 1/\left(\left|\sqrt{2}\Delta J\right|\right)$. For SLIC with DQ LAC, a complete interconversion of populations occurs if one chooses a longer duration $\tau_{SLIC}^{DQ} = 1/(|\Delta J|)$. Note that the populations oscillate back and forth when the LAC conditions are fulfilled, so that one returns to the initial state if one doubles the duration of the RF irradiation.

### *Symmetry of the density matrix*

Figure S3 shows a matrix representation of dimensions 64x64 of the density operator after triple SLIC excitation and application of a $T_{00}$ filter in the basis defined by Eqs. (S5) and (S7), which is very close to the eigenbasis.

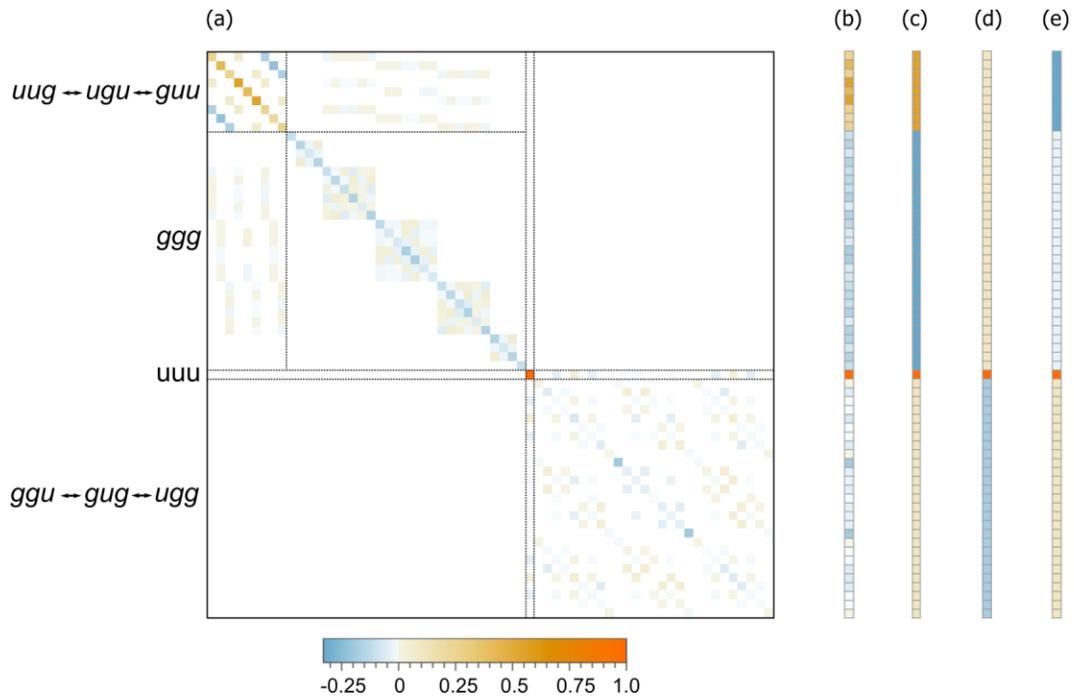

Figure S3. (a) Matrix representation of the 64x64 density operator in the basis defined by Eqs. (S5) and (S7) of the 6 protons of the $AA'MM'XX'$ system after triple SLIC to convert the magnetization into LLS. A $T_{00}$ filter was incorporated into the simulations to eliminate terms with ranks and coherence orders different from zero. The remaining off-diagonal elements represent zero-quantum coherences. The local permutation symmetries of the states are indicated in shorthand on the left side. (b) Vector composed of the diagonal elements of matrix shown in (a). (c) Vector representing the superposition of the three two-spin product terms $-1/3\left(\hat{I}^A \cdot \hat{I}^{A'} + \hat{I}^M \cdot \hat{I}^{M'} + \hat{I}^X \cdot \hat{I}^{X'}\right)$. (d) Vector representing the superposition of the three four-spin product terms $-1/3\left(\hat{I}^A \cdot \hat{I}^{A'}\hat{I}^M \cdot \hat{I}^{M'} + \hat{I}^A \cdot \hat{I}^{A'}\hat{I}^X \cdot \hat{I}^{X'} + \hat{I}^M \cdot \hat{I}^{M'}\hat{I}^X \cdot \hat{I}^{X'}\right)$. (e) Vector representing the unique six-spin product term $-\left(\hat{I}^A \cdot \hat{I}^{A'}\hat{I}^M \cdot \hat{I}^{M'}\hat{I}^X \cdot \hat{I}^{X'}\right)$. Note that the operators shown in (c), (d), and (e) do not have any off-diagonal elements in the considered basis.

States that are over-populated with respect to demagnetized states are indicated in orange, under-populated states in blue. The color scale, which extends from -1/4 to +1, was normalized with respect to 1, the population of the $S_0^{AA'} S_0^{MM'} S_0^{XX'}$ state. Note the population imbalances between the overpopulated orange (*guu, ugu,* and *uug*) states and underpopulated blue (*ggg*) triplet states within the globally *gerade* manifold (top left block), and the imbalance between the orange *uuu* and the blue (*ugg, gug, ggu*) states within the globally *ungerade* manifold (bottom right block).

All simulations were performed using the Spin Dynamica software package that allows one to solve the Liouville - von Neumann equations. Only unitary transformations of the density operator under coherent terms in the Hamiltonian have been considered, without any dissipative effects, i.e., without relaxation. The full relaxation superoperator can be derived from the knowledge of all dipole-dipole interactions and relevant spectral densities to account for both internal and overall motions, to estimate the lifetimes of all terms by diagonalization of the Liouvillian. This is beyond the scope of this work. Experimental results (see Figure 4 in the main text) show that the decays are mono-exponential.

### *Quantum yields of the excitation of LLS*

Numerical simulations show that the long-lived components of density matrix created after poly-SLIC can be represented by Eq. (7) in the main text, where $\hat{\sigma}_{LLS}$ is represented by a linear combination of normalised orthogonal operators, so that values of the $\lambda_i$ coefficients can be compared with each other. Depending on the excitation scheme (we distinguish 9 methods in the main text), one obtains different values for the coefficients $\lambda_{AA'}$, $\lambda_{MM'}$, $\lambda_{XX'}$, $\lambda_{AA'MM'}$, $\lambda_{MM'XX'}$, $\lambda_{AA'XX'}$, $\lambda_{AA'MM'XX'}$, as shown in Table S1.

Polychromatic Excitation of Delocalized Long-Lived Proton Spin States in Aliphatic Chains

Table S1. Calculated coefficients $\lambda_i$ of the seven product operator terms in Eq. (7) of the main text, excited using 9 different poly- and monochromatic SLIC methods applied to a chain with $n = 3$ aliphatic CH$_2$ groups making up an $AA'MM'XX'$ system. The coefficients represent the efficiency (quantum yield) of LLS excitation and were calculated as $\mathrm{Tr}\{(\sigma_{LLS}/\|\hat{I}_z\|)^\dagger (\hat{A}/\|\hat{A}\|)\}$, where $\|\hat{A}\| = \sqrt{\mathrm{Tr}\{\hat{A}^\dagger\hat{A}\}}$ and $\hat{I}_z = \hat{I}_z^A + \hat{I}_z^{A'} + \hat{I}_z^M + \hat{I}_z^{M'} + \hat{I}_z^X + \hat{I}_z^{X'}$ is the total initial magnetization of all 6 spins. The optimum RF amplitudes $\nu_{RF}$ are expressed as multiples of the intrapair scalar couplings $|J_{intra}| = 14$ Hz. The optimum durations of various SLIC irradiation schemes are $\tau_{SLIC}^{SQ} \approx 1/(|\sqrt{2}\Delta J|)$ and $\tau_{SLIC}^{DQ} \approx 1/(|\Delta J|)$. Double-quantum (DQ) LAC excitation requires less RF amplitude but takes more time than LLS excitation by single-quantum (SQ) LAC. The coefficients $\lambda_i$ correspond to projections of $\hat{\sigma}_{LLS}$ onto the operator $i$ and are expressed in percent. A value of 100% means that the initial magnetization $\hat{I}_z$ would be entirely transformed into the corresponding operator term.

| Method | Irradiation frequencies | Optimum RF amplitude | Optimum duration | $\lambda_A$ | $\lambda_M$ | $\lambda_X$ | $\lambda_{AM}$ | $\lambda_{AX}$ | $\lambda_{MX}$ | $\lambda_{AMX}$ |
|---|---|---|---|---|---|---|---|---|---|---|
| SQ single SLIC | $\nu_A$ | $2|J_{intra}|$ | $\tau_{SLIC}^{SQ}$ | 11 | 6.3 | 4.7 | -10 | -2.2 | 0.4 | -0.5 |
| SQ single SLIC | $\nu_M$ | $2|J_{intra}|$ | $\tau_{SLIC}^{SQ}$ | 7.2 | 14 | 7.2 | -6.2 | 4.0 | -6.2 | -4.6 |
| SQ single SLIC | $\nu_X$ | $2|J_{intra}|$ | $\tau_{SLIC}^{SQ}$ | 4.7 | 6.3 | 11 | 0.4 | -2.2 | -10 | -0.5 |
| SQ double SLIC | $\nu_A, \nu_M$ | $2|J_{intra}|$ | $\tau_{SLIC}^{SQ}$ | 3.7 | 7.5 | 10 | -1.5 | -3.3 | -8.5 | 0.6 |
| SQ double SLIC | $\nu_M, \nu_X$ | $2|J_{intra}|$ | $\tau_{SLIC}^{SQ}$ | 10 | 7.5 | 3.7 | -8.5 | -3.3 | -1.5 | 0.6 |
| SQ double SLIC | $\nu_A, \nu_X$ | $2|J_{intra}|$ | $\tau_{SLIC}^{SQ}$ | 7.7 | 16 | 7.7 | -6.6 | 4.5 | -6.6 | -5.3 |
| DQ double SLIC | $\nu_A, \nu_M$ | $|J_{intra}|$ | $\tau_{SLIC}^{DQ}$ | 14 | 13 | 0.6 | -15 | -5.1 | -4.7 | 5.2 |
| DQ double SLIC | $\nu_M, \nu_X$ | $|J_{intra}|$ | $\tau_{SLIC}^{DQ}$ | 0.6 | 13 | 14 | -4.7 | -5.1 | -15.4 | 5.2 |
| DQ triple SLIC | $\nu_A, \nu_M, \nu_X$ | $|J_{intra}|$ | $\tau_{SLIC}^{DQ}$ | 19 | 8.5 | 19 | -6.6 | -12 | -6.6 | -0.8 |

*Quantum yields of the reconversion of LLS*

The quantum yields of the reconversion of LLS shown in Tables S2 and S3 were calculated in analogy to the excitation yields in Table S1.

Table S2. Reconversion efficiencies starting with a normalized sum of three two-spin product terms $\hat{\mathbf{I}}^A \cdot \hat{\mathbf{I}}^{A'} + \hat{\mathbf{I}}^M \cdot \hat{\mathbf{I}}^{M'} + \hat{\mathbf{I}}^X \cdot \hat{\mathbf{I}}^{X'}$ in Eq (7) of the main text with equal coefficients $\lambda_{AA'} = \lambda_{MM'} = \lambda_{XX'} = 1/3$. The optimum durations of various SLIC irradiation schemes are $\tau_{SLIC}^{SQ} \approx 1/\left(\left|\sqrt{2}\Delta J\right|\right)$ and $\tau_{SLIC}^{DQ} \approx 1/(|\Delta J|)$. The resulting magnetization of the three pairs $(\hat{I}_x^A + \hat{I}_x^{A'})$, $(\hat{I}_x^M + \hat{I}_x^{M'})$ and $(\hat{I}_x^X + \hat{I}_x^{X'})$ is expressed in % of the amplitude they would have after a $\pi/2$ pulse applied to thermal equilibrium, thus providing a measure of the quantum yield of the reconversion. Note that single and double SLIC methods only yield magnetization of 2 or 4 spins, respectively.

| Method | Irradiation frequencies | Optimum RF amplitude | Optimum duration | $\hat{I}_x^A + \hat{I}_x^{A'}$ | $\hat{I}_x^M + \hat{I}_x^{M'}$ | $\hat{I}_x^X + \hat{I}_x^{X'}$ |
|---|---|---|---|---|---|---|
| SQ single SLIC | $\nu_A$ | $2|J_{intra}|$ | $\tau_{SLIC}^{SQ}$ | 22 | 0 | 0 |
| SQ single SLIC | $\nu_M$ | $2|J_{intra}|$ | $\tau_{SLIC}^{SQ}$ | 0 | 29 | 0 |
| SQ single SLIC | $\nu_X$ | $2|J_{intra}|$ | $\tau_{SLIC}^{SQ}$ | 0 | 0 | 22 |
| SQ double SLIC | $\nu_A, \nu_M$ | $2|J_{intra}|$ | $\tau_{SLIC}^{SQ}$ | 14 | 6.9 | 0 |
| SQ double SLIC | $\nu_M, \nu_X$ | $2|J_{intra}|$ | $\tau_{SLIC}^{SQ}$ | 0 | 6.9 | 14 |
| SQ double SLIC | $\nu_A, \nu_X$ | $2|J_{intra}|$ | $\tau_{SLIC}^{SQ}$ | 16 | 0 | 16 |
| DQ double SLIC | $\nu_A, \nu_M$ | $|J_{intra}|$ | $\tau_{SLIC}^{DQ}$ | 15 | 13 | 0 |
| DQ double SLIC | $\nu_M, \nu_X$ | $|J_{intra}|$ | $\tau_{SLIC}^{DQ}$ | 0 | 13 | 15 |
| DQ triple SLIC | $\nu_A, \nu_M, \nu_X$ | $|J_{intra}|$ | $\tau_{SLIC}^{DQ}$ | 16 | 14 | 16 |

Polychromatic Excitation of Delocalized Long-Lived Proton Spin States in Aliphatic Chains

Table S3. Reconversion efficiencies starting with the sum of the three four-spin product terms $(\hat{I}^A \cdot \hat{I}^{A'})(\hat{I}^M \cdot \hat{I}^{M'}) + (\hat{I}^A \cdot \hat{I}^{A'})(\hat{I}^X \cdot \hat{I}^{X'}) + (\hat{I}^M \cdot \hat{I}^{M'})(\hat{I}^X \cdot \hat{I}^{X'})$ in Eq. (7) of the main text with equal coefficients $\lambda_{AA',MM'} = \lambda_{AA',XX'} = \lambda_{MM',XX'} = 1/3$. The optimum durations of various SLIC irradiation schemes are $\tau_{SLIC}^{SQ} \approx 1/(|\sqrt{2}\Delta J|)$ and $\tau_{SLIC}^{DQ} \approx 1/(|\Delta J|)$. The resulting magnetization of the three pairs $(\hat{I}_x^A + \hat{I}_x^{A'})$, $(\hat{I}_x^M + \hat{I}_x^{M'})$ and $(\hat{I}_x^X + \hat{I}_x^{X'})$ is expressed in % of the amplitude they would have after a $\pi/2$ pulse applied to thermal equilibrium, thus providing a measure of the quantum yield of the reconversion. Note that single and double SLIC methods can only yield magnetization of 2 or 4 of the 6 spins, respectively. Since the coefficients of the six-spin product term $(\hat{I}^A \cdot \hat{I}^{A'})(\hat{I}^M \cdot \hat{I}^{M'})(\hat{I}^X \cdot \hat{I}^{X'})$ in Table S1 are small, its reconversion yield has not been calculated.

| Method | Irradiation frequencies | Optimum RF amplitude | Optimum duration | $\hat{I}_x^A + \hat{I}_x^{A'}$ | $\hat{I}_x^M + \hat{I}_x^{M'}$ | $\hat{I}_x^X + \hat{I}_x^{X'}$ |
|---|---|---|---|---|---|---|
| SQ single SLIC | $\nu_A$ | $2|J_{intra}|$ | $\tau_{SLIC}^{SQ}$ | 12 | 0 | 0 |
| SQ single SLIC | $\nu_M$ | $2|J_{intra}|$ | $\tau_{SLIC}^{SQ}$ | 0 | 9 | 0 |
| SQ single SLIC | $\nu_X$ | $2|J_{intra}|$ | $\tau_{SLIC}^{SQ}$ | 0 | 0 | 12 |
| SQ double SLIC | $\nu_A, \nu_M$ | $2|J_{intra}|$ | $\tau_{SLIC}^{SQ}$ | 5.3 | 8 | 0 |
| SQ double SLIC | $\nu_M, \nu_X$ | $2|J_{intra}|$ | $\tau_{SLIC}^{SQ}$ | 0 | 8 | 5.3 |
| SQ double SLIC | $\nu_A, \nu_X$ | $2|J_{intra}|$ | $\tau_{SLIC}^{SQ}$ | 4.4 | 0 | 4.4 |
| DQ double SLIC | $\nu_A, \nu_M$ | $|J_{intra}|$ | $\tau_{SLIC}^{DQ}$ | 13 | 12 | 0 |
| DQ double SLIC | $\nu_M, \nu_X$ | $|J_{intra}|$ | $\tau_{SLIC}^{DQ}$ | 0 | 12 | 13 |
| DQ triple SLIC | $\nu_A, \nu_M, \nu_X$ | $|J_{intra}|$ | $\tau_{SLIC}^{DQ}$ | 4.4 | 0 | 4.4 |

### *Scalar couplings of the AA'MM'XX' system of DSS*

Table S4. Scalar *J*-couplings [Hz] of NaSO$_3$CH$_2$CH$_2$CH$_2$Si(CH$_3$)$_3$ (2,2-Dimethyl-2-silapentane-5-sulfonate sodium salt, or DSS, dissolved in D$_2$O), resulting from the analysis of the multiplets in a conventional 500 MHz $^1$H NMR spectrum using ANATOLIA, and chemical shifts $\nu_A$, $\nu_M$ and $\nu_X$ [Hz] with respect to the signal of the methyl groups of DSS.

| | |
|---|---|
| $\nu_A$ | 1453 |
| $\nu_M$ | 879 |
| $\nu_X$ | 314 |
| $J_{AA'}$ | -14.1 |
| $J_{MM'}$ | -13.8 |
| $J_{XX'}$ | -14.5 |
| $J_{intra} = (J_{AA'} + J_{MM'} + J_{XX'})/3$ | -14 |
| $J_{AM} = J_{A'M'}$ | 10.6 |
| $J_{A'M} = J_{AM'}$ | 5.3 |
| $J_{MX} = J_{M'X'}$ | 11.8 |
| $J_{M'X} = J_{MX'}$ | 5.2 |
| $\langle \Delta J \rangle = [(J_{AM} - J_{AM'}) + (J_{MX} - J_{MX'})]/2$ | 6 |
| $J_{AX} = J_{A'X'} = J_{AX'} = J_{A'X}$ | -0.2 |

### Titration of DSS with bovine serum albumin (BSA)

To evaluate the potential of delocalized LLS in aliphatic chains for drug screening, titration experiments were performed to determine the contrast $C_{LLS}$ between free DSS and DSS that is weakly interacting with the protein bovine serum albumin (BSA) (Figure S4). The addition of a small amount of protein, so that the concentration ratio is [BSA]/[DSS] ≈ $10^{-4}$ leads to an abrupt drop of the observed lifetime $T_{LLS}$ by almost factor of 2. The effective concentration and rotational correlation time of BSA may be affected by its propensity to adhere to the surface of the glass tube.

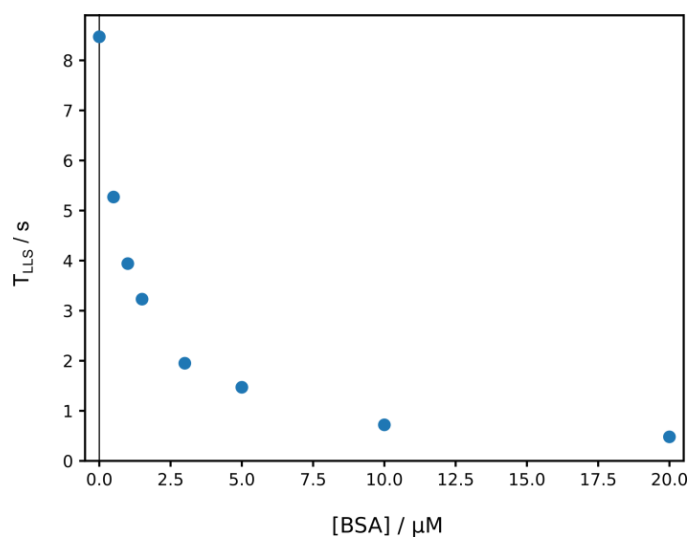

Figure S4. Dependence of the lifetime $T_{LLS}$ of the delocalized LLS excited and reconverted with single SLIC applied at the chemical shift $v_A = v_{A'}$ of DSS with an RF amplitude $2|J_{intra}| = 28$ Hz and a duration $\tau_{SLIC}^{SQ} = 110$ ms. The concentration [DSS] = 5 mM was constant, while the concentration of the protein was titrated over a range $0 \leq [BSA] \leq 20$ µM. The first two points correspond to [BSA] = 0 and 0.5 µM.

Conventional NMR approaches to drug screening monitor the effects of binding on relaxation times $T_1$, $T_{1\rho}$ and $T_2$ (line-broadening.) These methods only show evidence of binding when the concentration of BSA was approximately 100 times higher, i.e., when [BSA]/[DSS] > $10^{-2}$. The interaction between DSS to BSA was confirmed by observing broadening of the $^1$H NMR lines from *ca.* 1 Hz to 1.6 Hz for [DSS] = 5 mM and [BSA] = 20 µM. This suggests that long-lived states in aliphatic chains may significantly reduce the concentrations of proteins required for drug screening. In 2D Nuclear Overhauser Effect Spectroscopy (NOESY), negative cross-peaks are expected for small molecules such as DSS in the absence of BSA. Upon addition of BSA, the cross-peaks become positive because the averaged rotational correlation time of DSS is significantly decreased by its interaction with BSA. We have indeed observed that the signs of the cross-peaks between all $CH_2$ and $CH_3$ groups of free DSS were positive (with respect to the positive diagonal peaks) for [DSS] = 50 mM and [BSA] = 50 µM.

### *Experimental details*

All experiments were conducted using a Bruker NEO spectrometer with a static field $B_0 = 11.9$ T (proton Larmor frequency $\nu_0 = 500$ MHz) with a 10 mm broadband double resonance BBO probe. The *RF* inhomogeneity leads to an intensity loss of 16 % after a 2π pulse. The durations of the SLIC pulses were optimized experimentally to $\tau_{SLIC}^{SQ} = 110$ ms and $\tau_{SLIC}^{DQ} = 156$ ms for DQ LAC. The *RF* amplitudes were optimised to $\nu_{RF} = 14$ Hz (for DQ LAC) or 28 Hz (for SQ LAC). Superpositions of two or three simultaneous selective pulses were generated by phase modulation. The recovery delay between subsequent experiments was set to 60 s to allow the LLS to decay completely. The initial $(\pi/2)_x$ proton pulses of 25 µs duration had an *RF*-field amplitude of 10 kHz. All spin simulations were obtained with Spin Dynamica [1]. The *J*-couplings in a conventional $^1$H NMR spectrum were determined with ANATOLIA [2]. All experiments except those in Figure S4 were performed at 300 K on samples containing 33 mM DSS (Sigma Aldrich) dissolved in $D_2O$ without degassing and without buffer. The titration experiments shown in Figure S4 were obtained at 800 MHz and 298 K with [DSS] = 5 mM, 0 < [BSA] < 20 µM in 15 mM phosphate buffer in $D_2O$ at pH = 7. Each point shown in Figure S4 was acquired using the sequence of Figure 1 of the main text with 8 different delays $\tau_{rel}$, requiring *ca*. 13 min per point. The $T_{00}$ filter [3] consisted of three non-selective π/2 pulses and three pulsed field gradients to eliminate terms with ranks $l = 1$ and 2 and coherence orders $p = 1$ and 2. A 4-step phase alternation of the SLIC pulses along the ±y axes with concomitant addition and subtraction of the signals was used to suppress undesirable pathways [4]. The zero-quantum coherences were not suppressed.

### References

[1] C. Bengs and M. H. Levitt, *SpinDynamica: Symbolic and Numerical Magnetic Resonance in a Mathematica Environment*, Magn. Reson. Chem. **56**, 374 (2018).

[2] D. A. Cheshkov, K. F. Sheberstov, D. O. Sinitsyn, and V. A. Chertkov, *ANATOLIA: NMR software for spectral analysis of total lineshape*, Magn. Reson. Chem. **56**, 449 (2018).

[3] M. C. D. Tayler, *Filters for Long-Lived Spin Order*, in *Long-Lived Nuclear Spin Order* (The Royal Society of Chemistry, 2020), *Chapter 10,* pp. 188–208.

[4] A. N. Pravdivtsev, A. S. Kiryutin, A. V. Yurkovskaya, H.-M. Vieth, and K. L. Ivanov, *Robust Conversion of Singlet Spin Order in Coupled Spin-1/2 Pairs by Adiabatically Ramped RF-Fields*, J. Magn. Reson. **273**, 56 (2016).